\documentclass[12pt,twoside]{article}   
\usepackage[super,sort,comma]{natbib}
\usepackage{amsfonts}
\usepackage{amsmath}
\usepackage{multirow}

\usepackage{fancyhdr}		

\usepackage[section]{placeins}   %

\usepackage{graphicx}

\makeatletter \renewcommand\@biblabel[1]{$^{#1}$} \makeatother
\newcommand{\cen}[1]{\begin{center} #1 \end{center}}

\usepackage[mathlines]{lineno}

\usepackage{hyperref}
\hypersetup{ colorlinks,
    citecolor=blue,
    filecolor=blue,
    linkcolor=blue,
    urlcolor=blue
}

\usepackage{xcolor}

\definecolor{gray}{rgb}{0.6,0.6,0.6}
\definecolor{red}{rgb}{0.85,0,0}
\definecolor{green}{rgb}{0,0.85,0}
\definecolor{blue}{rgb}{0,0,0.85}
\definecolor{beige}{rgb}{0.92,0.87,0.78}
\usepackage[all]{hypcap}    

\begin{document}

\cen{\sf {\Large {\bfseries \textcolor{blue}{Fluence} Adaptation for Task-based Dose Optimization in X-ray Phase-Contrast Imaging} \\  
\vspace*{10mm}
Chengpeng Wu$^{1,2}$, Yuxiang Xing$^{1,2}$, Li Zhang$^{1,2,*}$, Zhiqiang Chen$^{1,2}$, Xiaohua Zhu$^{1,2}$, Xi Zhang$^{3}$ and Hewei Gao$^{1,2,*}$} \\
1. Department of Engineering Physics, Tsinghua University, Haidian District, Beijing, China\\
2. Key Laboratory of Particle \& Radiation Imaging (Tsinghua University) of Ministry of Education, Haidian District, Beijing, China\\
3. Department of Radiology, Fudan University Shanghai Cancer Center, 270 Dongan Road, Shanghai 200032,
China. \\
}

\pagenumbering{roman}
\setcounter{page}{1}
\pagestyle{plain}
\cen{Corresponding authors: Li Zhang, Hewei Gao. \\
Email: zli@mail.tsinghua.edu.cn, hwgao@tsinghua.edu.cn}

\begin{abstract}
\noindent {\bf Purpose:} X-ray phase-contrast imaging (XPCI) can provide multiple contrasts with great potentials for clinical and industrial applications, including conventional attenuation, \textcolor{blue}{phase contrast and dark field}. Grating-based imaging (GBI) and edge-illumination (EI) are two promising types of XPCI as the conventional x-ray sources can be directly utilized. For GBI and EI systems, the phase-stepping acquisition with multiple exposures at a constant \textcolor{blue}{fluence} is usually adopted in the literature. This work, however, attempts to challenge such a constant \textcolor{blue}{fluence} concept during the phase-stepping process and proposes a \textcolor{blue}{fluence} adaptation mechanism for dose reduction. \\
{\bf Method:} Given the importance of patient radiation dose for clinical applications, numerous studies have tried to reduce patient dose in XPCI by \textcolor{blue}{altering imaging system designs}, data acquisition and information retrieval. Recently, analytic multi-order moment analysis has been proposed to improve the computing efficiency. In these algorithms, \textcolor{blue}{multiple} contrasts can be calculated by summing together the weighted phase-stepping curves (PSCs) with some kernel functions\textcolor{blue}{, which} suggests us that the raw data at different steps have different contributions for the \textcolor{blue}{noise in retrieved contrasts}. Therefore, it is possible to improve the noise performance by adjusting the \textcolor{blue}{fluence} distribution during the phase-stepping process directly. Based on analytic retrieval formulas and the \textcolor{blue}{Gaussian} noise model for detected signals, we \textcolor{blue}{derived an optimal adaptive} \textcolor{blue}{fluence} distribution, which is proportional to the absolute weighting kernel functions and the root of original sample PSCs acquired under the constant \textcolor{blue}{fluence}. \textcolor{blue}{Considering that the original sample PSC might be unavailable, we proposed two practical forms for GBI and EI systems, which are also able to reduce the contrast noise when comparing with the constant fluence distribution.} Since the kernel functions are target contrast and information retrieval algorithm dependent, our proposed \textcolor{blue}{fluence} adaptation mechanism provides a way of realizing a task-based dose optimization while keeping the same noise level. \\
{\bf Results:} To validate our analyses, \textcolor{blue}{simulations and experiments are conducted for} GBI and EI systems. Simulated results demonstrate that the dose reduction \textcolor{blue}{ratio between our proposed fluence distributions and the typical constant one can be about 20\% for the phase contrast}, which is consistent with our theoretical predictions. \textcolor{blue}{Although the experimental noise reduction ratios are a little smaller than the theoretical ones, synthetic and real experiments both observe better noise performance by our proposed method.} Our simulated results also give out the effective ranges of the parameters of the PSCs, such as the visibility in GBI, the standard deviation and the mean value in EI, providing a guidance for the use of our proposed approach in practice.\\
{\bf Conclusions:} In this paper, we propose a \textcolor{blue}{fluence} adaptation mechanism for task-based dose optimization in XPCI, which can be applied to GBI and EI systems. Our proposed method explores a new direction for dose reduction, and may also be further extended to other types of XPCI systems and information retrieval algorithms.\\

\noindent Key words: x-ray phase-contrast imaging, grating-based imaging, fluence adaptation, dose optimization
\end{abstract}

\newpage     

\tableofcontents

\newpage

\setlength{\baselineskip}{0.7cm}      

\pagenumbering{arabic}
\setcounter{page}{1}
\pagestyle{fancy}
\section{INTRODUCTION}\label{sec:intro}
X-ray phase contrast imaging (XPCI) is a promising imaging technology that can provide informative phase contrast in addition to the conventional attenuation contrast. 
 Since 1960s, many researchers have attempted to visualize the phase contrast by various approaches, which can be mainly classified into the crystal-based interferometry (CBI)\cite{bonse1965x}, the analyzer-based imaging (ABI)\cite{chapman1997diffraction}, the propagation-based imaging (PBI)\cite{jacobsen1990xray}, the grating-based imaging (GBI)\cite{david2002differential,momose2003demonstration,pfeiffer2006phase,huang2009alternative}, and the edge-illumination (EI)\cite{olivo2001an,olivo2007a}. All methods above can obtain the phase information of a scanned object, but in different forms such as the original phase shift, the first-order differential phase shift and the second-order differential phase shift. According to the principles of these methods, the first three (CBI, ABI, and PBI) are more feasible for x-ray sources with strong coherence like synchrotron sources, while the last two (GBI and EI) can directly work with conventional x-ray sources. Therefore, GBI and EI have great potentials for wide industrial and medical applications. In the last decade, numerous studies have discussed about the system design, information retrieval algorithms, data acquisition, and various applications of GBI and EI. Besides, Pfeiffer et al. \cite{pfeiffer2008hard} introduced a new scattering-sensitive contrast into this field, i.e., the dark-field, which has been demonstrated to be a good representation for many diseases, especially related to the lung and bone joints. Considering the valuable meanings in practice, GBI and EI are going from the laboratory to the hospital, and a couple of prototypes have been established for clinical trials recently\cite{astolfo2017large,willer2018x,arboleda2019towards, zhang2019human,li2020a}.

For clinical applications, the patient radiation dose is one of the most important indices, and has impeded the clinical development of XPCI for many years. In this respect, both GBI and EI face quite a big challenge in the use of phase-stepping (PS) approach commonly, which requires one of the gratings or coded apertures to move along the transverse direction perpendicular to the grating lines step by step over one pitch, and therefore needs multiple exposures at one scanning angle. Initially, the PS approach was developed in visible-light interferometry \cite{creath1988v}, and it was then adapted to the early GBI system (i.e., the Talbot interferometry \cite{momose2003demonstration}) by Weitkamp et al. \cite{weitkamp2005x}. Using the PS approach, one can acquire a so-called phase-stepping curve (PSC) at each detector pixel (also called as the illumination curve in EI). To eliminate the inherent system effects, one needs to firstly obtain a \textcolor{blue}{flat field PSC} (without the sample) and then a sample PSC, from both of which typical contrasts (such as the attenuation, phase-contrast and dark-field) can be retrieved using an information retrieval algorithm.

During the phase-stepping, it is a routine to adopt a constant \textcolor{blue}{fluence} at all steps, which \textcolor{blue}{can be easily carried out}. To the best of our knowledge, \textcolor{blue}{there is lack of literature focusing on the effects of step-varying fluence} distributions during the phase-stepping acquisition. Inspired by recent analytic \textcolor{blue}{multi-order moment analyses} \cite{modregger2018direct,wu2019trigonometric}, we have found that the raw PSCs will be weighted across steps in the post-processing information retrieval. Intuitively, the weighted PSCs have different noise contributions compared with the original acquired PSCs, which suggests that there may be some rooms for further noise optimization. On the other hand, the weighting process also means that we can change the noise performance of retrieved contrasts directly by adjusting the \textcolor{blue}{fluence} distribution during phase-stepping. With these thoughts, in this paper, we attempt to \textcolor{blue}{explore a step-varying fluence} paradigm in the PS process. Therefore, among all \textcolor{blue}{non-constant fluence} distributions, we develop a \textcolor{blue}{fluence} adaptation method for task-based noise optimization in GBI and EI. The ``task-based'' qualifier means that the proposed optimization should be designed according to the pre-determined contrast (i.e., the phase-contrast or the dark-field) and its information retrieval algorithm. Of course, with the task-based \textcolor{blue}{fluence} adaptation method, one can reduce radiation dose while maintaining the same image noise level from the typical \textcolor{blue}{fluence} acquisition scheme. 

In the following Section \ref{sec:methods}, we will briefly review the information retrieval algorithms, and derive the task-based dose optimization method \textcolor{blue}{with step-varying fluence distributions} theoretically. In Section \ref{sec:results}, \textcolor{blue}{numerical simulation, synthetic and real} experimental results are \textcolor{blue}{presented} to validate our approach. Section \ref{sec:discu} includes some discussions followed by a brief conclusion in Section \ref{sec:concl}.

\section{MATERIALS AND METHODS}\label{sec:methods}
\subsection{\label{sec:method1}The Information Retrieval Algorithms}
The information retrieval is one of the most important processes to obtain desired contrasts, and has been actively investigated in GBI and EI. So far, there are mainly two trends among them. The first one is the classic Fourier component analysis (FCA), which retrieves typical contrasts by computing harmonic coefficients of both the \textcolor{blue}{flat field PSC} and the sample PSC. FCA is simple to understand and efficient to implement, but has the strict cosinusoidal assumption for both PSCs and the inherent phase-wrapping problem. The other one is the later developed multi-order moment analysis (MMA). Totally different from FCA, MMA regards typical contrasts as moments of the underlying small-angle x-ray scattering (SAXS) distribution \cite{modregger2012imaging}. Based on the convolution relationship between PSCs and SAXS distributions \cite{wang2009quantitative}, MMA obtains the SAXS distribution by a deconvolution algorithm, and then retrieves multiple contrasts by computing moments of the SAXS distribution. Compared with the FCA, MMA can obtain additional higher moments useful in some applications \cite{modregger2012imaging, modregger2017interpretation}. Simultaneously, MMA no longer needs the cosinusoidal assumption and therefore avoids the annoying phase-wrapping problem naturally. However, the critical deconvolution procedure in MMA is usually done iteratively like the Lucy-Richardson deconvolution and therefore, is quite time-consuming and difficult to determine a proper stopping criteria during the iteration. 

To overcome these limitations of the original deconvolution-based MMA (named DB-MMA in this paper), some analytic MMA methods are developed recently \cite{modregger2018direct,wu2019trigonometric}. The first one is referred as the direct MMA (D-MMA) \cite{modregger2018direct} that is applicable to EI and ABI at first, and it is extended to GBI with some pre-processing operations\cite{zhu2019direct}. Then a generalized D-MMA (GD-MMA) is developed by using the trigonometric orthogonality of PSCs \cite{wu2019trigonometric}, which can be utilized to improve the noise performance of the original D-MMA. The typical contrasts retrieved by D-MMA and GD-MMA can be expressed as,
\begin{eqnarray}
\label{equ:G1} && \textcolor{blue}{A\rightarrow M_0(g) =\int_{-\pi}^{\pi} g(\phi) d\phi= \frac{M_0(s)}{M_0(f)} = \frac{\int_{-\pi}^{\pi} s(\phi) d\phi}{\int_{-\pi}^{\pi} f(\phi) d\phi},}\\
\label{equ:G2} && \textcolor{blue}{P\rightarrow \bar{M}_1(g) =\frac{M_1(g)}{M_0(g)}= \frac{\int_{-\pi}^{\pi}\phi \cdot g(\phi) d\phi}{\int_{-\pi}^{\pi}g(\phi) d\phi}= \bar{M}_1(s) - \bar{M}_1(f) }\\ \nonumber
&&\quad \quad \quad \quad \quad \textcolor{blue}{= \frac{\int_{-\pi}^{\pi}h_1(\phi) \cdot s(\phi) d\phi}{\int_{-\pi}^{\pi}s(\phi) d\phi} - \frac{\int_{-\pi}^{\pi}h_1(\phi) \cdot f(\phi) d\phi}{\int_{-\pi}^{\pi}f(\phi) d\phi},}\\
\label{equ:G3} && \textcolor{blue}{D\rightarrow \tilde{M}_2(g) =\frac{\int_{-\pi}^{\pi}\left[\phi-\bar{M}_1(g)\right]^2\cdot g(\phi) d\phi}{M_0(g)}=\frac{M_2(g)}{M_0(g)}-\bar{M}_1^2(g) = \tilde{M}_2(s) - \tilde{M}_2(f) }\\ \nonumber
&&\textcolor{blue}{\quad \quad \quad \quad \quad = \frac{\int_{-\pi}^{\pi}h_2(\phi)s(\phi) d\phi}{\int_{-\pi}^{\pi}s(\phi) d\phi} - \frac{\int_{-\pi}^{\pi}h_2(\phi)f(\phi) d\phi}{\int_{-\pi}^{\pi}f(\phi) d\phi} - \bar{M}_1^2(s) + \bar{M}_1^2(f),}
\end{eqnarray}
\textcolor{blue}{where, $A$, $P$ and $D$ denote the attenuation, phase-contrast and dark-field signals, respectively; $M_n(g)$ denotes the $n$th-order un-normalized and un-centralized moments of $g(\phi)$; $\bar{M}_n(g)$ denotes the corresponding $n$th-order normalized but un-centralized moments; $\tilde{M}_n(g)$ denotes the corresponding $n$th-order normalized and centralized moments;} $\phi\in [-\pi, \pi]$ is the lateral offset of PSCs; $s(\phi), f(\phi), g(\phi)$ denote the sample PSC, the \textcolor{blue}{flat field PSC} and the SAXS distribution, respectively; \textcolor{blue}{$h_1(\phi)$ and $h_2(\phi)$ are the kernel functions corresponding to the phase-contrast and the dark-field.} In the original D-MMA\cite{modregger2018direct}, the kernel functions are $h_1(\phi)=\phi$ and $h_2(\phi)=\phi^2$, while $h_1(\phi)=2\sin(\phi)$ and $h_2(\phi)=\pi^2/3-4\cos(\phi)$ in the noise-optimal GD-MMA for GBI\cite{wu2019trigonometric}.

From the analytic expressions in D-MMA and GD-MMA above, it is found that multiple contrasts can be calculated by summing together the weighted \textcolor{blue}{PSCs} with some kernel functions. \textcolor{blue}{These non-uniform weighting} kernel functions suggest that the raw data across the steps has different contributions for retrieved contrasts and so do their corresponding noises. Therefore, it is possible to adjust the \textcolor{blue}{fluence distribution in phase-stepping} to improve the noise performance directly, or reduce patient dose at the same noise level equivalently. Specifically, in this study a \textcolor{blue}{fluence} adaption mechanism is developed for a task-based dose optimization as follows.

\subsection{\label{sec:method2}\textcolor{blue}{Fluence} Adaptation for Task-based Dose Optimization}
According to Eqs. (\ref{equ:G2}) and (\ref{equ:G3}) in D-MMA and GD-MMA, \textcolor{blue}{the normalized first-order moment and normalized second-order moment retrieval formulas in the presence of noise, corresponding to the phase-contrast and the dark-field, can be unified as:}
\begin{equation}
\label{equ:1}
C=\frac{\int{h(\phi)\left[\bar{s}(\phi)+n_s(\phi)\right]d\phi}}{\int{\left[\bar{s}(\phi)+n_s(\phi)\right]d\phi}} - \frac{\int{h(\phi)\bar{f}(\phi)d\phi}}{\int{\bar{f}(\phi)d\phi}},
\end{equation}
where, $C$ denotes the target \textcolor{blue}{signal} and $\phi$ denotes the phase term of PSCs; $\bar{s}(\phi)$ and $\bar{f}(\phi)$ denote the noise-free sample PSC and the \textcolor{blue}{flat field PSC} correspondingly acquired with high \textcolor{blue}{fluence}; $n_s(\phi)$ denotes the statistical noises in the sample PSC, while the \textcolor{blue}{flat field PSC} can be considered as noise-free acquired under high \textcolor{blue}{fluence} without any actual doses for the scanning sample; $h(\phi)$ denotes the weighting kernel function, for example, $h(\phi)=\phi$ for the phase-contrast in D-MMA. Using Eq. (\ref{equ:1}) above, a task-based dose optimization can be \textcolor{blue}{carried out for XPCI systems.}

As described in Section \ref{sec:intro}, our optimization goal in this study is to reduce the total \textcolor{blue}{absorption dose of the sample in data acquisition while keeping the same noise variance of a target signal}. Vice versa, the equivalent goal is to minimize the noise level among various \textcolor{blue}{fluence} distributions under the same total \textcolor{blue}{absorption dose}. Without loss of generality, we assume the signals of acquired sample PSCs are proportional to the normalized \textcolor{blue}{fluence} distribution applied during the phase stepping data acquisition, i.e.,
\begin{equation}
\label{equ:2}
\bar{s}(\phi) = \bar{s}_0(\phi) \cdot t(\phi),
\end{equation}
where, $\bar{s}_0(\phi)$ denotes the original sample PSC under the constant \textcolor{blue}{fluence}, and $t(\phi)$ denotes the normalized \textcolor{blue}{fluence} distribution. In general, the \textcolor{blue}{fluence} at every step can be adjusted by controlling the exposure time or the tube current, both of which are proportional to the \textcolor{blue}{fluence}.

In practice, the PS process is discrete, with a limited number of steps, $N$. Thus, the equivalent optimization problem can be written mathematically as:
\begin{eqnarray}
\label{equ:3} &&\textcolor{blue}{\text{minimize}_t \text{Var}\left(\frac{\sum_{i=1}^N h(i)\left[\bar{s}_0(i)t(i)+n_s(i)\right]/t(i)}{\sum_{i=1}^N \left[\bar{s}_0(i)t(i)+n_s(i)\right]/t(i)}\right) }\\
\label{equ:4} && \textcolor{blue}{\text{s.t.} \sum_{i=1}^N t(i)= N,  \qquad \qquad i=1,2,...,N,}
\end{eqnarray}
where, \textcolor{blue}{$\text{Var}(\cdot)$ denotes the variance of a statistic variable. Both the numerator and denominator in Eq. (\ref{equ:3}) are divided by $t(i)$ to keep the signal value consistent with the original target signal $C$. }

\textcolor{blue}{After some simplifications based on the noise property and sample PSCs in Appendix A}, the optimization problem defined in Eqs. (\ref{equ:3}) and (\ref{equ:4}) can be rewritten as:
\begin{eqnarray}
\label{equ:10} &&\text{minimize}_t \sum_{i=1}^N \frac{h^{2}(i)\bar{s}_0(i)}{t(i)} \\
\label{equ:11} && \text{s.t.} \sum_{i=1}^N t(i)= N.
\end{eqnarray}

Next, we will solve the problem with the classic Lagrange multiplier method. First, the Lagrange function can be written as:
\begin{equation}
\label{equ:12} L(t,\lambda) = \sum_{i=1}^N \frac{h^2(i)\bar{s}_0(i)}{t(i)} + \lambda\left(\sum_{i=1}^N t(i) - N\right),
\end{equation}
where $\lambda$ is a constant.
Then let us set the partial derivatives of the Lagrange function to $t(i)$ and $\lambda$ as zero, i.e.,
\begin{eqnarray}
\label{equ:13} &&\frac{\partial L}{\partial t(i)}=-\frac{h^2(i)\bar{s}_0(i)}{t^2(i)}+\lambda=0,\\
\label{equ:14} &&\frac{\partial L}{\partial \lambda}=\sum_{i=1}^N t(i) - N=0.
\end{eqnarray}

From Eq. (\ref{equ:13}), it can be seen that an optimal adaptive \textcolor{blue}{fluence} distribution $t^*(i)$ is proportional to the absolute value of the weighting function $h(i)$ multiplied by the root of the original sample PSC $\bar{s}_0(i)$, i.e.,
\begin{equation}
\label{equ:15} t^*(i)=\sqrt{\frac{h^2(i)\bar{s}_0(i)}{\lambda}}\propto |h(i)|\sqrt{\bar{s}_0(i)}.
\end{equation}

Substituting Eq. (\ref{equ:15}) into Eq. (\ref{equ:14}), one gets the optimal adaptive \textcolor{blue}{fluence} distribution as,
\begin{equation}
\label{equ:160} t^*(i)= |h(i)|\sqrt{\bar{s}_0(i)} \cdot \frac{N}{\sum_{i=1}^N |h(i)|\sqrt{\bar{s}_0(i)}}.
\end{equation}

Until here, it is derived that Eq. (\ref{equ:160}) is the ideal solution of the optimization problem defined in Eqs. (\ref{equ:3}) and (\ref{equ:4}) for the GBI and EI systems. \textcolor{blue}{Equation (\ref{equ:160}) provides a fundamental guidance on how to apply our proposed method in practice. The prerequisite of using Eq. (\ref{equ:160}) is that the kernel function and the original sample PSC are known. The former can be determined by the target contrast and the selected information retrieval algorithm, while the latter needs to be obtained by a pre-scan under low dose or acquired prior images. In terms of this prior, our proposed method shares some similarities with some reconstruction algorithms for low dose data\cite{chen2008Prior}, which requires obtaining prior images to help improve image quality. While, here we use the prior to guide the design of x-ray fluence during the phase-stepping scan. With the XPCI technology from the bench to the bed, it can be expected that this type of prior images may be available in the near future, due to regular repeated physical examinations.}

\subsection{\label{sec:method3}\textcolor{blue}{Two Practical Adaptive Fluence Distribution Forms}}

As most GBI and EI systems are at laboratories rather than in hospitals at present, the sample PSC $\bar{s}_0(i)$ with the constant \textcolor{blue}{fluence} cannot be obtained in advance sometimes. \textcolor{blue}{To achieve a comparable performance with our proposed method in Eq. (\ref{equ:160}), we propose two practical forms that can be directly applied without the known sample PSC.} 

\textcolor{blue}{Firstly, for samples with relatively weak refraction, the original sample PSCs acquired with a constant fluence} will have relatively small shifts and deformations compared with the original \textcolor{blue}{flat field PSC}. Therefore, one can roughly approximate the ideal solution for both GBI and EI systems in Eq. (\ref{equ:160}) by replacing $\bar{s}_0(i)$ with $\bar{f}_0(i)$, which  can be available before the scanning of samples, i.e.,
\begin{equation}
\label{equ:16} t^*_1(i) \approx |h(i)|\sqrt{\bar{f}_0(i)} \cdot \frac{N}{\sum_{i=1}^N |h(i)|\sqrt{\bar{f}_0(i)}}.
\end{equation}

In particular for GBI,  \textcolor{blue}{considering a smaller difference (see about 25\%) caused by the visibility of the sample PSC, the weighting flux difference across steps is mainly determined by the kernel functions (varying from 0\% to 200\%), and therefore, the adaptive fluence distribution can be further approximated to be simply proportional to the absolute kernel functions, i.e.,}

\begin{equation}
\label{equ:17} t^*_2(i)\approx |h(i)| \cdot  \frac{ N}{\sum_{i=1}^N |h(i)|}.
\end{equation}

It should be clarified that the kernel function $h(\phi)$ is target contrast and information retrieval algorithm dependent, such as $h_1^{D-MMA}(\phi)=\phi$ and $h_2^{D-MMA}(\phi)=\phi^2$, while $h_1^{GD-MMA}(\phi)=2\sin(\phi)$ and $h_2^{GD-MMA}(\phi)=\pi^2/3-4\cos(\phi)$. Therefore, optimal adaptive \textcolor{blue}{fluence} distributions might be different for different target contrasts and adopted information retrieval algorithms. \textcolor{blue}{In this work, the adaptive distributions for D-MMA and GD-MMA are referred as Adap-D and Adap-GD, respectively. The approximation error depends on the phase-contrast and dark-field signals, which is analyzed further in Appendix B.}

\begin{figure}[htbp]
	\centering
	\includegraphics[width=0.65\columnwidth]{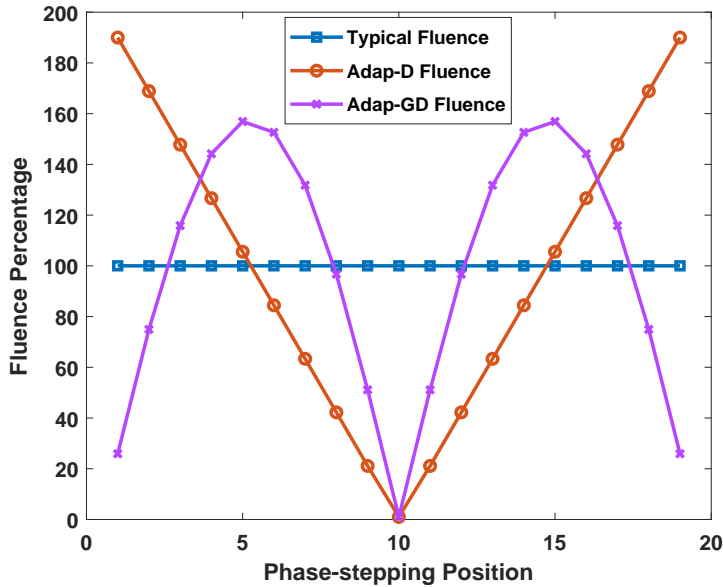}%
	\caption{\label{fig:1}The typical constant \textcolor{blue}{fluence} distribution and our proposed Type-2 practical adaptive ones in Eq. (\ref{equ:17}) for retrieving the phase-contrast by D-MMA and GD-MMA, respectively. The number of steps in phase-stepping is $N=19$ \textcolor{blue}{and the sample phase $\varphi_1$ in Eq. (\ref{equ:5}) is assumed to be zero.}}%
\end{figure}

\textcolor{blue}{In summary, our work proposes an optimal adaptive fluence design method in Eq. (\ref{equ:160}) theoretically. Moreover, limited by the actual imaging conditions, two practical forms in Eq. (\ref{equ:16}) called as the Type-1 form and Eq. (\ref{equ:17}) called as the Type-2 form are given. They are effective with no sample PSC priors for GBI systems, and the former is also applicable for EI systems.} \textcolor{blue}{To show our proposed method, according to the simplest form in Eq. (\ref{equ:17}), we calculate an adaptive fluence distribution for retrieving the phase-contrast in GBI using the D-MMA and GD-MMA algorithms, respectively, as shown in Fig. \ref{fig:1}.}

~
\subsection{\label{sec:method4}\textcolor{blue}{Validation Method}}
\textcolor{blue}{In order to validate our analyses, we compared the differential phase-contrast (DPC) and the dark-field contrast (DFC) retrieved by the D-MMA and GD-MMA algorithms on simulated data, synthetic experimental data, and real experimental data, respectively. The detailed parameters of simulated and real system setups are listed in Table \ref{tab:0}.} 

\textcolor{blue}{The simulation of GBI systems was conducted on our laboratory's simulation platform \cite{zhang2014sensitivity} for a Talbot-Lau system. The phantom used in the simulation is a polymethyl methacrylate (PMMA) cylinder of 10 mm in diameter, which is built with simple numerical ellipses. In order to explore the effective range of our proposed adaptive \textcolor{blue}{fluence} distributions, we chose a series of visibilities to generate the cosine PSCs as Eq. (\ref{equ:5}) and compared their noise performance.  For EI systems, we also validated our proposed method by using the simulated Gaussian distributions as Eq. (\ref{equ:18}) and changing the parameters (the standard deviation and the mean value) to find their effective ranges for reference in practice. To evaluate noise performance, we added various levels of noises to the simulated data. We repeated  $10^6$ times for every level of noise to eliminate the random effect.}

\begin{figure}[htbp]
	\begin{center}
		\includegraphics[width=0.75\columnwidth]{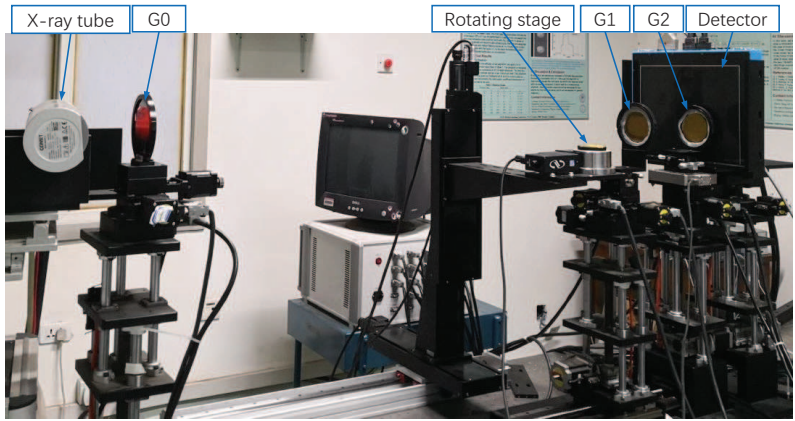}
		\caption{\label{fig:0} The experimental Talbot-Lau platform in our laboratory including three gratings.}
	\end{center}
\end{figure}

\textcolor{blue}{The synthetic experimental data was originally acquired in a previous study \cite{wu2019trigonometric}, which was weighted by the calculated adaptive fluence distributions according to our proposed method. For utilizing all data of original PSCs, we set the lowest adaptive fluence percentage as 1\% rather than 0 shown in the Fig. \ref{fig:1}. In this part, we utilized two sets of data, both of which are breast tissue specimens. One was acquired in a Talbot system at the Shanghai Synchrotron Radiation Facility (SSRF), and the other was acquired in the Talbot-Lau system at our laboratory in Tsinghua University (shown in Fig. \ref{fig:0}). These specimens were originated from the right breast of a 51-year-old female patient from Fudan University Shanghai Cancer Center (FUSCC) in Shanghai, China. The noise levels in original PSCs are relatively low, so we also inserted noises to these synthetic experimental data for evaluating noise performances.}

\textcolor{blue}{To directly validate the effectiveness of our proposed method, we carried out a real experiment at our laboratory's Talbot-Lau system. The tube current and voltage were 24 mA and 35 kV, with the beam filtration of 0.8 mm Be inherent in our tube. For the typical constant acquisition, exposure time was set as 2.4 s for each step in the phase-stepping process, while exposure time in the adaptive fluence acquisition was calculated according to our proposed weighting method in Eq. (\ref{equ:17}). The samples used in the real experiment include a PMMA cylinder and a POM triangular prism. The latter has a constant differential phase contrast in theory. For comparison, we choose two smooth areas as ROIs to evaluate the noise performance. To eliminate the structure information in the retrieved images, we also acquired a set of low dose data with 0.1 s exposure time, and subtracted the corresponding contrasts to obtain pure noise images.}

\begin{table}[htbp]
	\begin{center}
		\caption{\label{tab:0} The parameters of simulated and experimental systems.}
		\begin{tabular}{ccc}
			\hline \hline
			& Simulated System&Experimental System\\\hline
			Type&Talbot-Lau&Talbot-Lau\\ 
			Source&Plane Wave&Comet MXR-160HP\\
			Detector&CsI Scintillation&Dexela 1512\\
			Pixel Size&75 $\mu$m&75 $\mu$m\\
			G0 pitch&16.8 $\mu$m&4.8 $\mu$m\\
			G1 pitch&4.2 $\mu$m&4.8 $\mu$m\\
			G2 pitch&2.4 $\mu$m&4.8$\mu$m\\
			No. of steps & 12&12\\
\hline\hline
		\end{tabular}
	\end{center}
\end{table}


\section{RESULTS}\label{sec:results}
\textcolor{blue}{How much noises can be reduced by using our proposed fluence distribution compared with the typical one? To figure out the nominal values, we calculate the object functions of our adaptive distribution and the typical one in Eq. (\ref{equ:10}), and then summarize their values in Table \ref{tab:1}.} From Table \ref{tab:1}, it can be seen that the nominal noise variances can be reduced more than 20\% by our proposed adaptive \textcolor{blue}{fluence} distributions (about 25\% for D-MMA and 20\% for GD-MMA). From the derivations above, our \textcolor{blue}{fluence} adaptation method can be applied to retrieve the typical contrasts by D-MMA and GD-MMA directly for GBI and EI systems.
\begin{table}[htbp]
	\centering
	\caption{\label{tab:1}The object function values in Eq. (\ref{equ:10}) calculated for the D-MMA and GD-MMA algorithms under three \textcolor{blue}{fluence} distributions in Fig. \ref{fig:1}.}
	\begin{tabular}{cccccc}
		\hline\hline
		PS Steps&Algorithm& Typical & Adap-D & Adap-GD & Reduction\\
		\hline
		\multirow{2}{*}{N=9}&D-MMA &29.24&\textbf{21.66}&38.57&25.92\%\\
		&GD-MMA &18.00&20.93&\textbf{14.29}&20.61\%\\
		\hline
		\multirow{2}{*}{N=19}&D-MMA &62.33&\textbf{46.62}&110.36&25.20\%\\
		&GD-MMA &38.00&45.84&\textbf{30.66}&19.32\%\\
		\hline\hline
	\end{tabular}
\end{table}
\subsection{Numerical Simulation Results}\label{sec:result1}
\subsubsection{For the GBI systems}\label{sec:result11}
\begin{figure}[htbp]
	\centering
	\includegraphics[width=0.48\columnwidth]{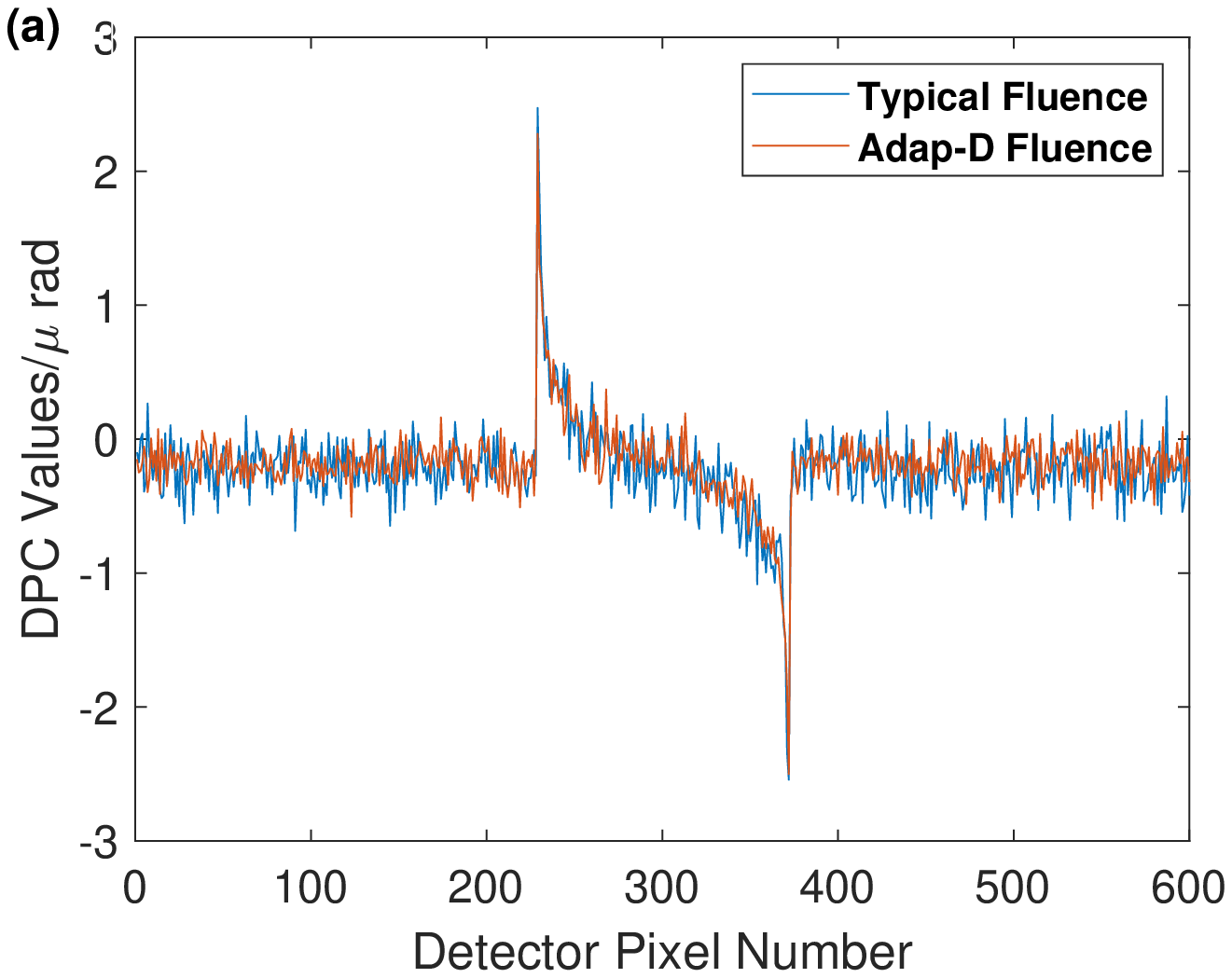}
	\includegraphics[width=0.48\columnwidth]{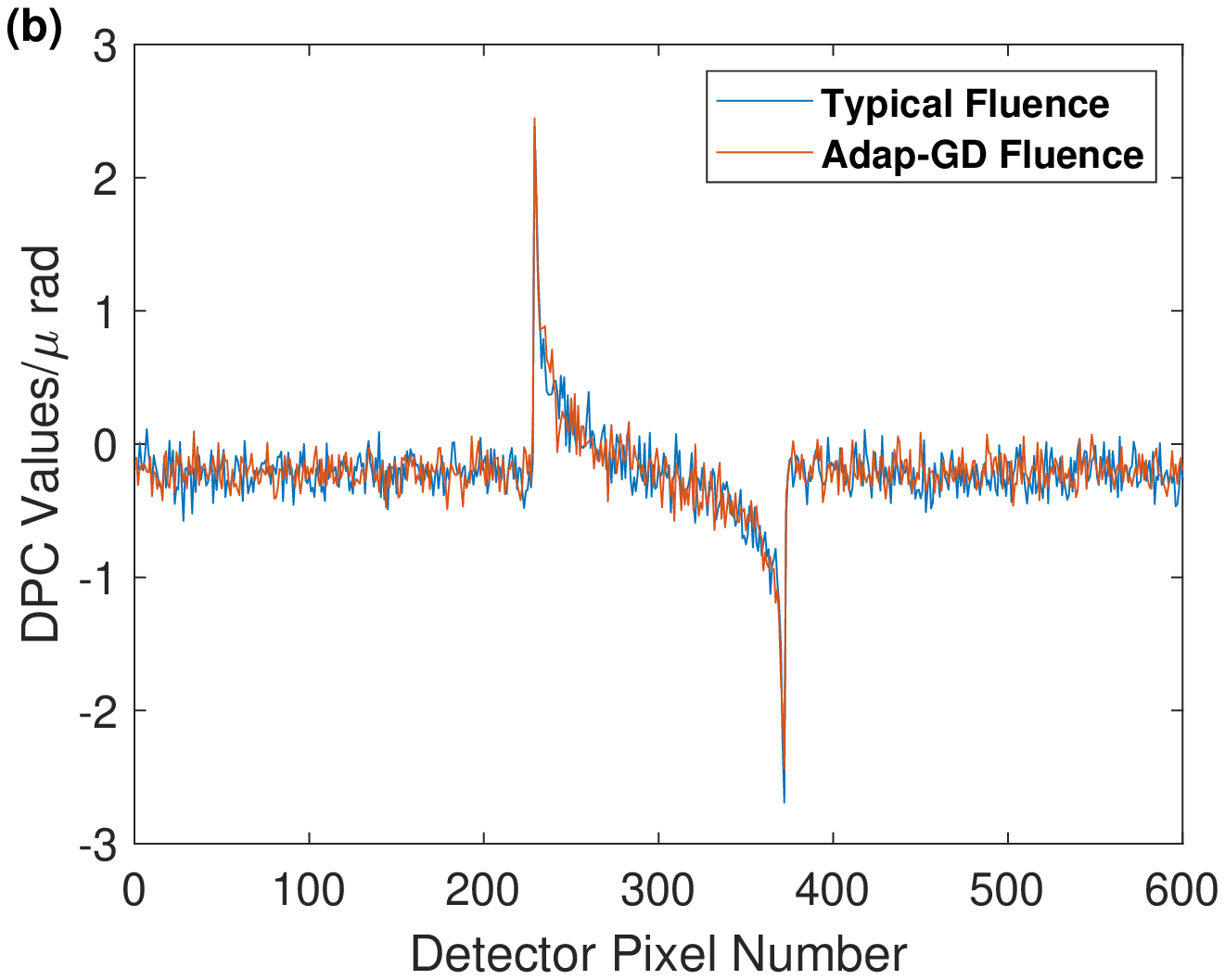}
	\caption{\label{fig:2}The simulated DPC profiles of a PMMA cylinder retrieved by the D-MMA (a) and GD-MMA (b) algorithms under the typical \textcolor{blue}{fluence} distribution and our proposed adaptive \textcolor{blue}{fluence} distributions.}%
\end{figure}

\begin{figure}[htbp]
	\centering
	\includegraphics[width=0.48\columnwidth]{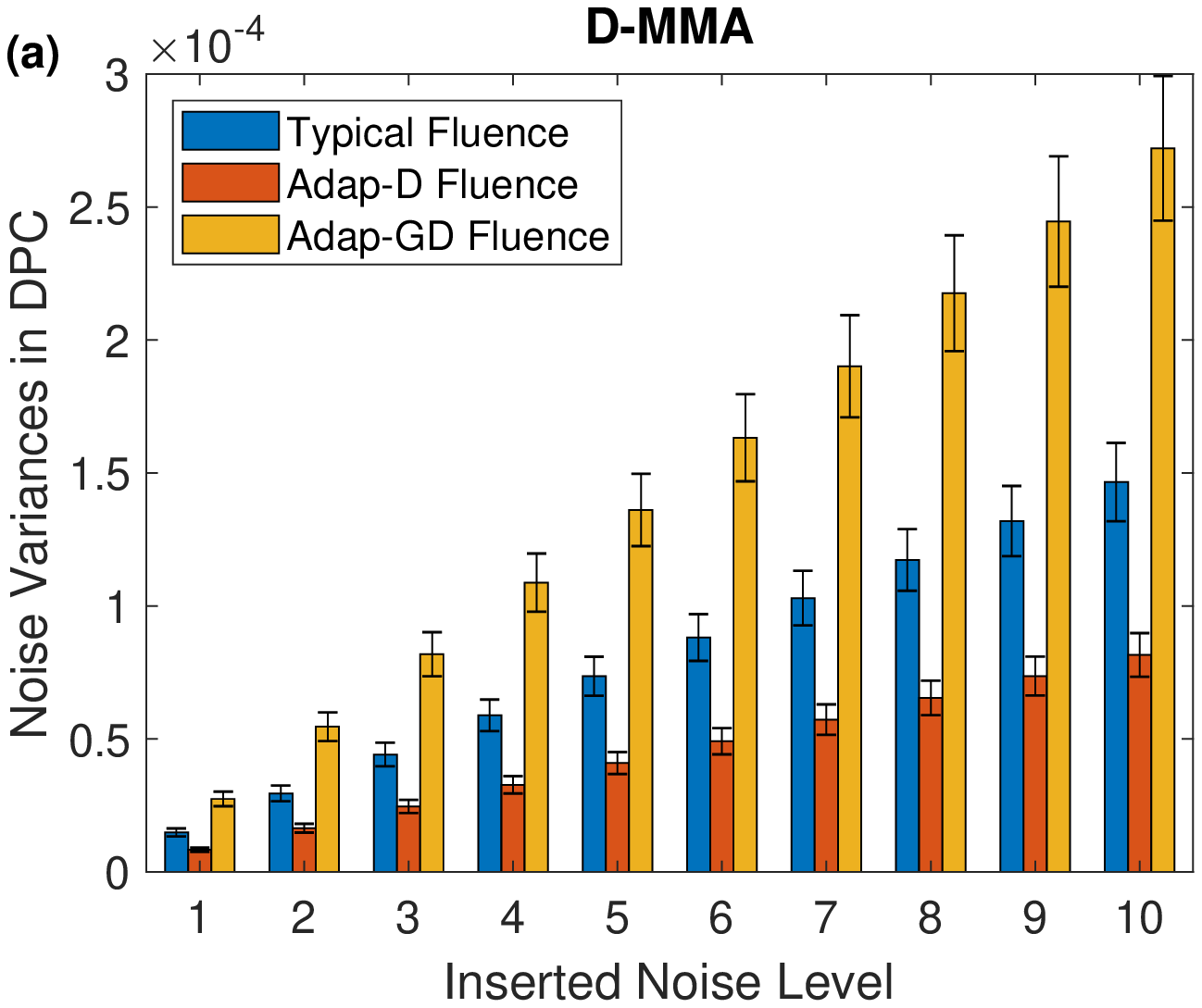}
	\includegraphics[width=0.48\columnwidth]{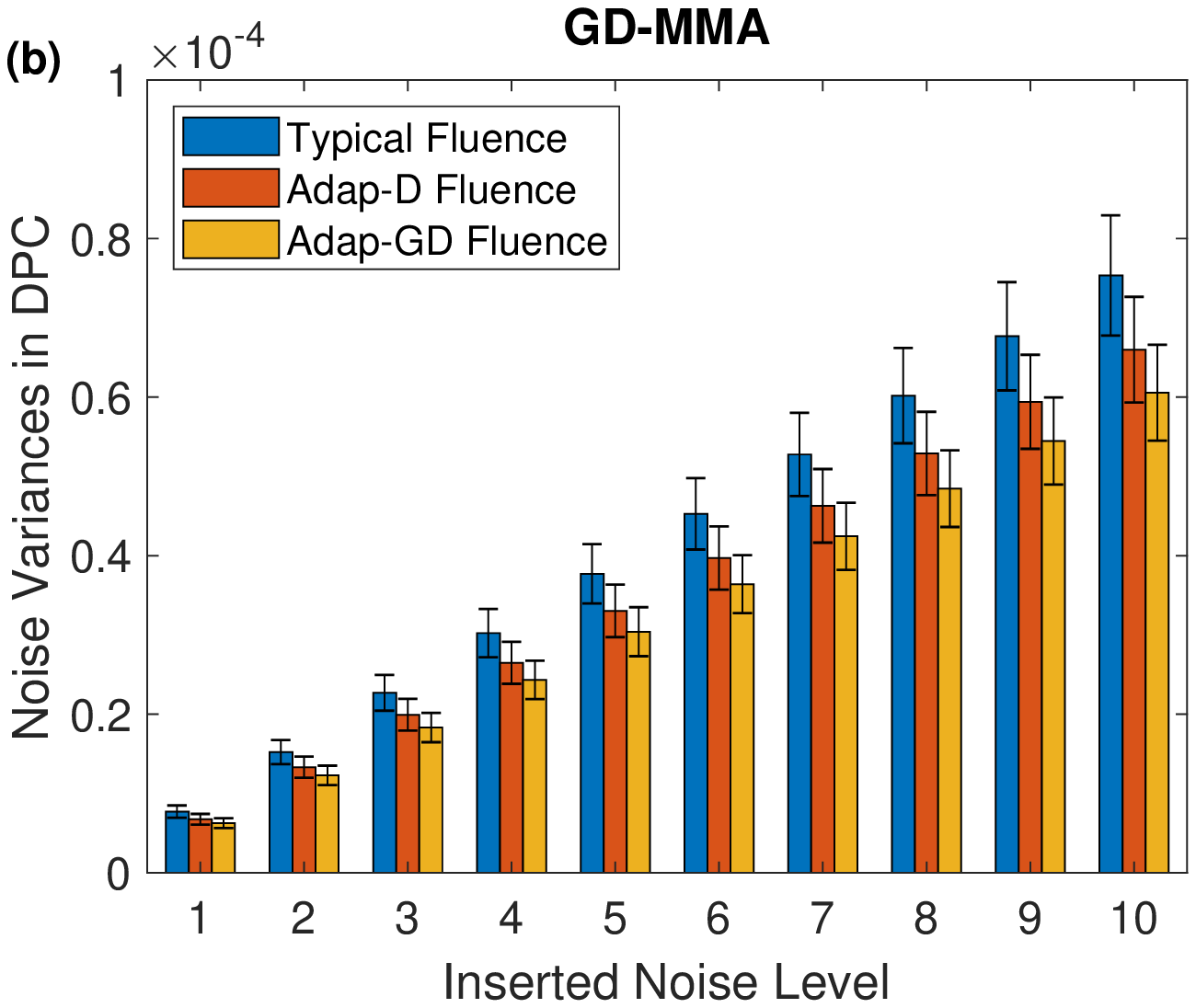}
	\caption{\label{fig:3}The simulated noise variances in DPC retrieved by the D-MMA (a) and GD-MMA (b) algorithms under three mentioned \textcolor{blue}{fluence} distributions and a range of inserted noise levels.}
\end{figure}

As shown in Fig. \ref{fig:2}, both the simulated DPC profiles of a PMMA cylinder retrieved by the D-MMA and GD-MMA algorithms under our proposed adaptive \textcolor{blue}{fluence} distributions in Eq. (\ref{equ:17}) have lower levels of noises, compared with that under the typical constant \textcolor{blue}{fluence} distribution. Further, as shown in Fig. \ref{fig:3}, the noise reduction ratios by using the proposed adaptive \textcolor{blue}{fluence} distribution are almost constant for various levels of inserted Gaussian noises.

~\\

\begin{figure}[htbp]
	\centering
	\includegraphics[width=0.45\columnwidth]{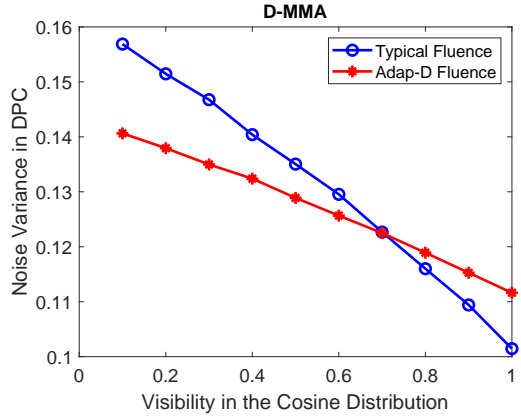}
	\caption{\label{fig:4}The simulated noise variances in DPC retrieved by the D-MMA algorithm under the typical \textcolor{blue}{fluence} distribution and our proposed adaptive \textcolor{blue}{fluence} distributions for various visibility of GBI systems.}
\end{figure}

Since there are some approximations in our derivations of Section \ref{sec:method3}, we try to explore the effective range of our proposed adaptive \textcolor{blue}{fluence} distributions in Eq. (\ref{equ:16}) and Eq. (\ref{equ:17}). As shown in Fig. \ref{fig:4}, the noise reduction ratio caused by our adaptive distributions decreases as the visibility increases, and the proposed \textcolor{blue}{fluence} distribution is better than the original one when the visibility is \textcolor{blue}{smaller than 70\%, which is mostly satisfied in practical systems.}

\subsubsection{For the EI systems}\label{sec:result12}
As explained in the Section \ref{sec:method4}, we simulated the raw PSCs in EI systems by generating some Gaussian curves within the range of $[-\pi, \pi]$ according to Eq. (\ref{equ:18}). It is obvious that there are two main parameters (the standard deviation $\sigma$ and the mean value
$\mu$) in the Gaussian model which may influence the noise performance. According to the literature\cite{olivo2007a,endrizzi2014absorption,zamir2017recent}, the collimated beam in EI usually causes symmetrical and narrow Gaussian-shape PSCs, which means the standard deviation $\sigma$ is small and the mean value $\mu$ is close to zero. 

\begin{figure}[htbp]
	\centering
	\includegraphics[width=0.45\columnwidth]{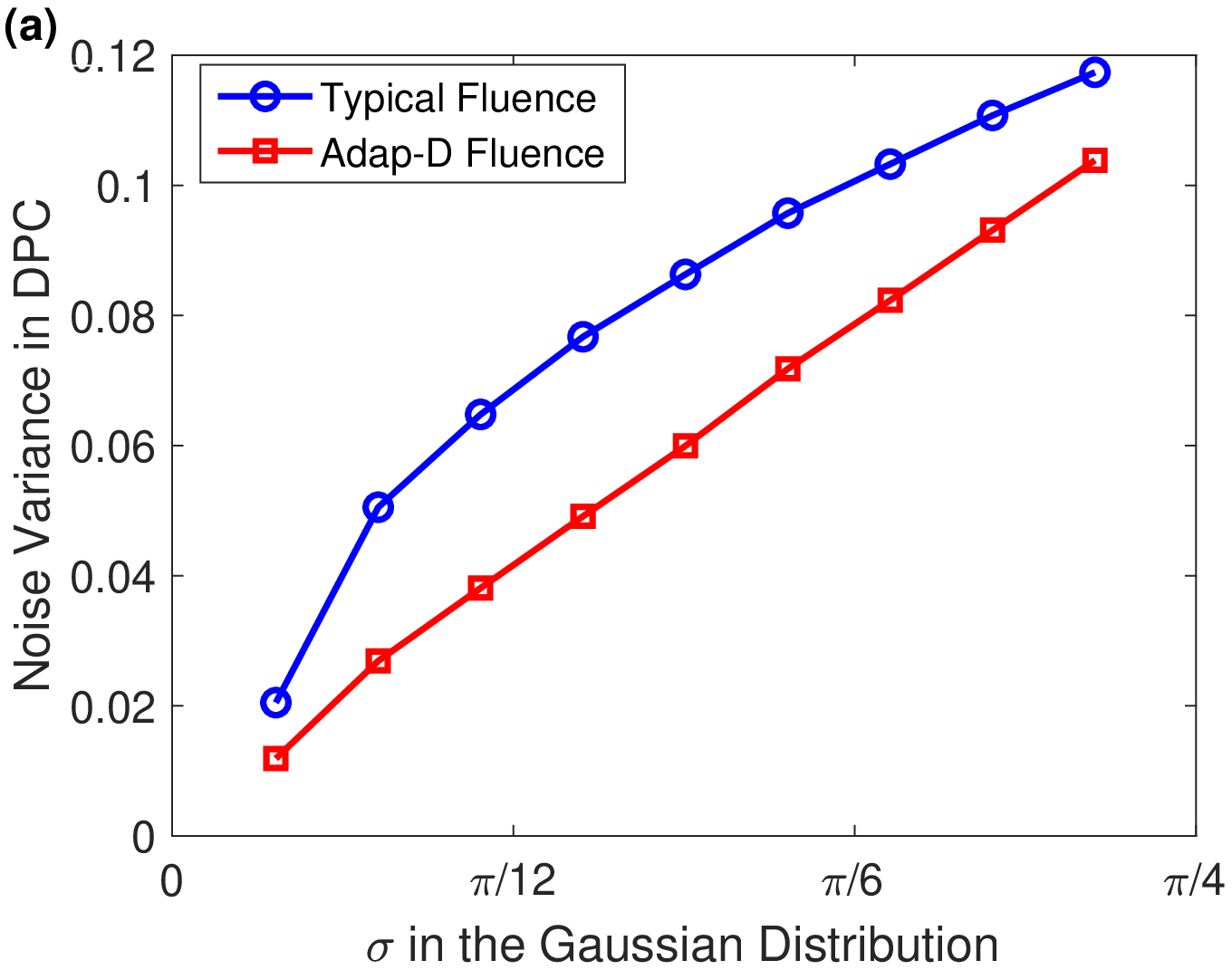}
	\includegraphics[width=0.45\columnwidth]{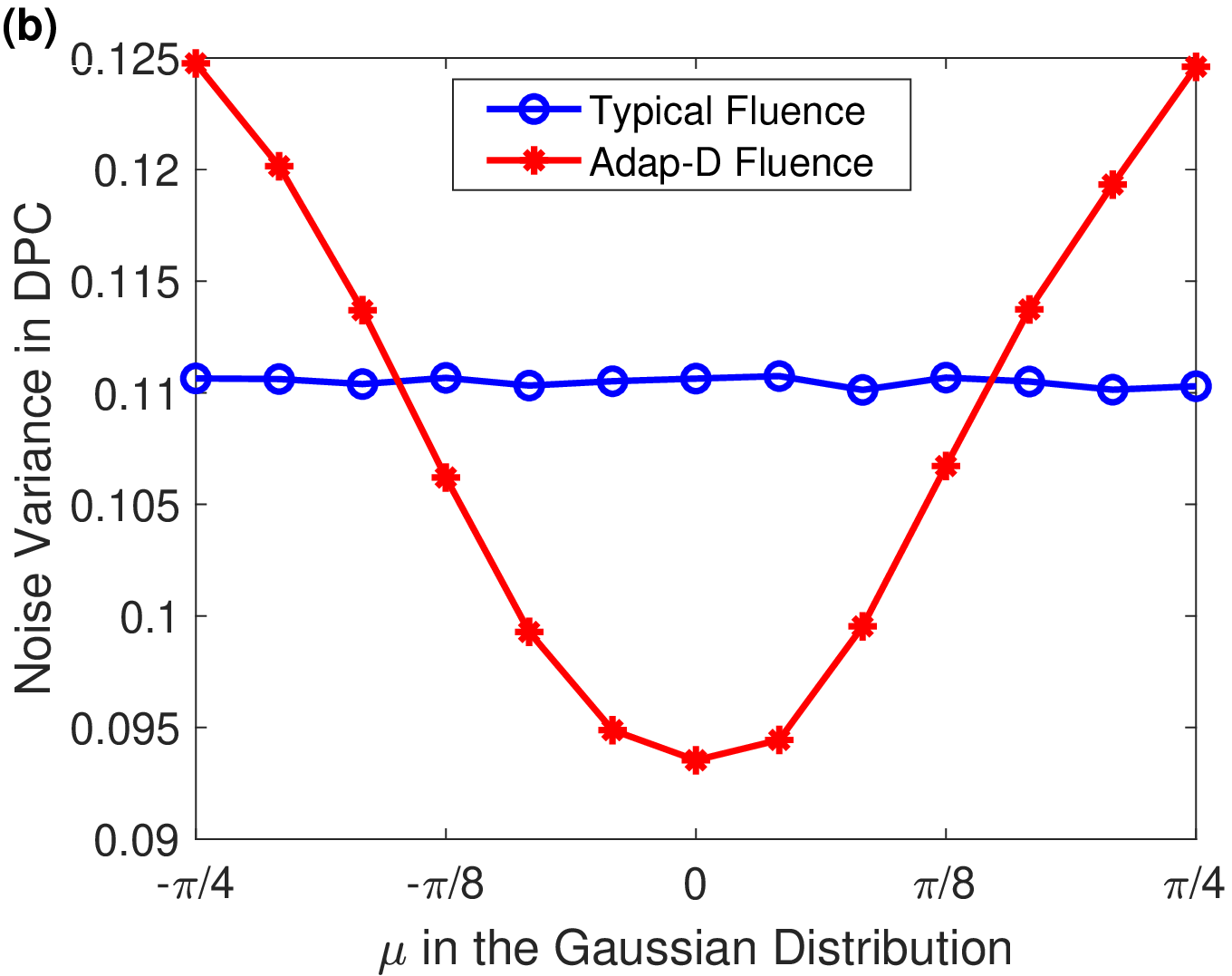}
	\caption{\label{fig:5}The simulated noise variances in DPC retrieved by D-MMA with the typical \textcolor{blue}{fluence} distribution and our proposed adaptive \textcolor{blue}{fluence} distributions for a range of standard deviations (a) and a range of mean values (b) in the PSCs of EI systems.}
\end{figure}

Similar with the GBI systems, we want to figure out the effective parameter ranges of our proposed adaptive \textcolor{blue}{fluence} distributions in Eq. (\ref{equ:16}) for the EI systems. In order to test one parameter at a time, the mean value is set as zero when changing the standard deviation, while the standard deviation is set as $\pi/6$ when changing the mean value. As shown in Fig. \ref{fig:5}(a), the adaptive \textcolor{blue}{fluence} method can reduce significant noises in the phase-contrast retrieved by the D-MMA algorithm when the standard deviation $\sigma$ of the Gaussian PSCs is smaller than $\pi/4$. Due to the discrete PS process, the PSCs will be truncated Gaussian distributions for EI systems and therefore, the ideal noise reduction ratio caused by our adaptive method will decrease as the $\sigma$ increases. But if the $\sigma$ is too small, most points in the PSCs will be close to zero. Therefore, the first point of both \textcolor{blue}{fluence} distribution in Fig. \ref{fig:5}(a) has very low noise variance. Also, from Fig. \ref{fig:5}(b), we can see that the effective range of our method for the mean value $\mu$ is about [$-\pi/8$, $\pi/8$]. Since we adopt the approximate adaptive \textcolor{blue}{fluence} distribution in Eq. (\ref{equ:16}) by substituting the original sample PSC with the \textcolor{blue}{flat field PSC} which is usually symmetrical, the noise reduction effect caused by our method will decrease as the absolute mean value of the sample PSC increases. These results can provide some references for the parameters of \textcolor{blue}{flat field PSCs} acquired in real EI systems when using our proposed method.

\subsection{\textcolor{blue}{Synthetic Experimental Results}}\label{sec:result2}
\begin{figure}[htbp]
	\centering
	\includegraphics[width=0.7\columnwidth]{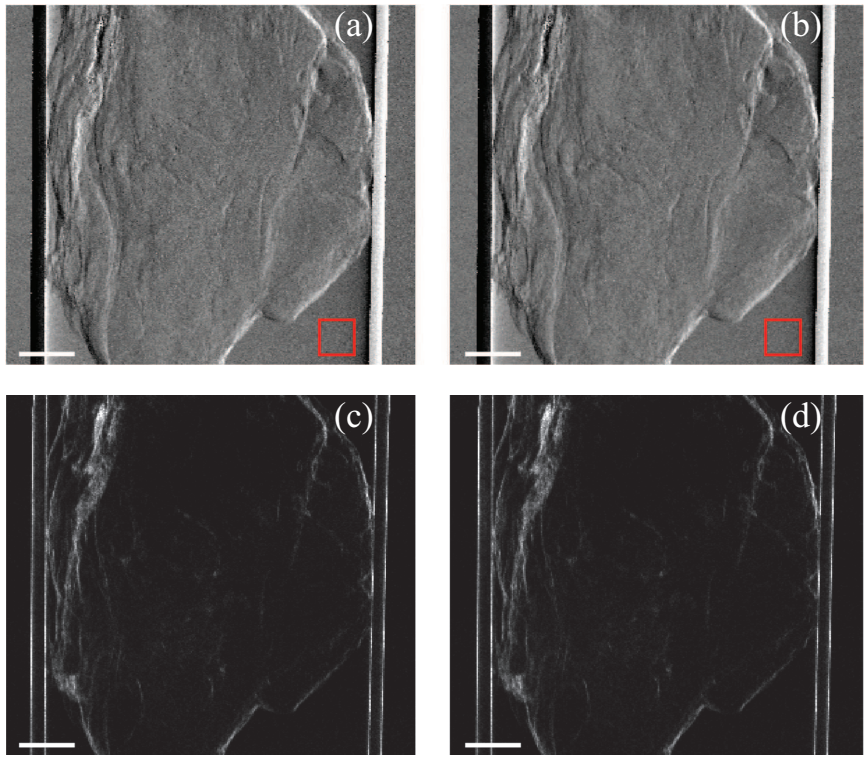}
	\caption{\label{fig:8}The \textcolor{blue}{synthetic} experimental DPC (a/b) and DFC (c/d) images of a breast specimen retrieved by the D-MMA algorithm under the typical \textcolor{blue}{fluence} distribution (a/c) and our proposed Type-1 adaptive \textcolor{blue}{fluence} distribution (b/d). This set of raw data was acquired on a Talbot-Lau system. The scale bar is 6 mm and the red box is the ROI area for evaluating the noise performance.}                                                  
\end{figure}

\begin{figure}[htbp]
	\centering
	\includegraphics[width=1\columnwidth]{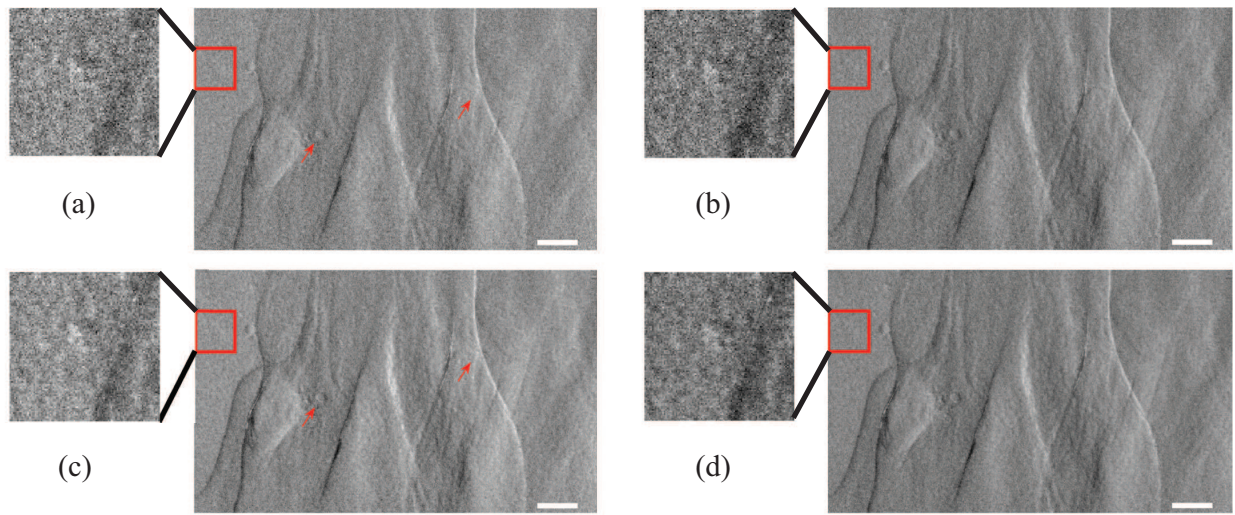}
	\caption{\label{fig:6}The \textcolor{blue}{synthetic} experimental DPC images of a breast specimen retrieved by the D-MMA (a/c) and GD-MMA (b/d) algorithms under the typical \textcolor{blue}{fluence} distribution (a/b) and our proposed Type-1 adaptive \textcolor{blue}{fluence} distribution(c/d), with the zoomed-in ROI area shown on the left. This set of raw data was acquired on a Talbot system. The scale bar is 6.5 mm and the red box is the ROI area for evaluating the noise performance.}                              
\end{figure}

\textcolor{blue}{The retrieved contrasts of synthetic experimental data by the D-MMA and GD-MMA algorithms are shown in Fig. \ref{fig:8} and Fig. \ref{fig:6}. Figure \ref{fig:8} presents DPC and DFC images retrieved by the D-MMA algorithm for a Talbot-Lau system, while Fig. \ref{fig:6} displays the DPC images retrieved by the D-MMA and GD-MMA algorithms for a Talbot system. For conciseness, we only show the DFC images in Fig. \ref{fig:8} and the images retrieved by GD-MMA in Fig. \ref{fig:6}. Both DPC and DFC images for the breast specimen in Fig. \ref{fig:8} indicate that structures are the same for different fluence distributions, but the image noise levels are different, especially for the DPC. In both Fig. \ref{fig:8} and Fig. \ref{fig:6}, sub-figures (a/c) and (b/d) are obtained with the typical constant fluence and our proposed Type-1 adaptive fluence, respectively. From the zoomed-in sub-images in Fig. \ref{fig:6}, it is clear to see that the \textcolor{blue}{fluence} adaptation strategy can reduce the noises both for the D-MMA and GD-MMA algorithms. Moreover, some structures (pointed by red arrows) in the retrieved image of Fig. \ref{fig:6} look much more clear by adopting our proposed adaptive fluence distribution.}

To compare the noises of different \textcolor{blue}{fluence} distributions quantitatively, the noise variances retrieved by the D-MMA and GD-MMA algorithms under three \textcolor{blue}{fluence} distributions are listed in Table \ref{tab:2} and Table \ref{tab:20}. These numbers demonstrate the effectiveness of our proposed adaptive \textcolor{blue}{fluence} distribution in terms of the noise reduction ratios. For the DPC, they are around 20\% in the Talbot system (18.96\% for D-MMA and 26.65\% for GD-MMA), which are consistent with our theoretical expectations in Table \ref{tab:1}. The noise reduction ratios in the Talbot-Lau system are a little lower (about 13\%), which is mainly caused by the higher system visibility (about 30\%) as explained in the simulated results in Fig. \ref{fig:4}. \textcolor{blue}{The ratios for the DFC are around 10\%, smaller than those for the DPC. While reducing the noise variances for both the DPC and DFC, their signal values are almost intact for all fluence distributions.}\\

\begin{table}[htbp]
	\centering
	\caption{\label{tab:2}The synthetic experimental noise variances ($10^{-6}$) of DPC in Fig. \ref{fig:8} and Fig. \ref{fig:6}, retrieved by D-MMA and GD-MMA under three \textcolor{blue}{fluence} distributions.}
	\begin{tabular}{ccccccc}
	\hline\hline
	System&Algorithm&Index&  Typical & Adap-D & Adap-GD & Reduction\\
	\hline
	\multirow{4}{*}{Talbot-Lau}&\multirow{2}{*}{D-MMA}&Signal &47.20&47.24&47.13&\\
	&&Noise&13.55&\textbf{11.53}&14.04&13.22\%\\
	&\multirow{2}{*}{GD-MMA}&Signal &47.94&47.82&47.46&\\
	&&Noise &11.57&11.92&\textbf{10.04}&13.18\%\\
	\hline
	\multirow{4}{*}{Talbot}&\multirow{2}{*}{D-MMA}&Signal &20.25&20.52&20.79&\\
	&&Noise &9.02&\textbf{7.31}&10.60&18.96\%\\
	&\multirow{2}{*}{GD-MMA}&Signal &20.88&20.03&20.98&\\
	&&Noise &6.11&7.32&\textbf{4.48}&26.65\%\\
	\hline\hline
	\end{tabular}
\end{table}

\begin{table}[htbp]
	\centering
	\caption{\label{tab:20}The \textcolor{blue}{synthetic experimental noise variances ($10^{-12}$) of DFC in Fig. \ref{fig:8} and Fig. \ref{fig:6}, retrieved by D-MMA and GD-MMA under three fluence distributions.}}
	\begin{tabular}{ccccccc}
		\hline\hline
		System&Algorithm&Index&  Typical & Adap-D & Adap-GD & Reduction\\
		\hline
		\multirow{4}{*}{Talbot-Lau}&\multirow{2}{*}{D-MMA}&Signal &12.70&12.48&12.12&\\
		&&Noise&8.37&\textbf{7.56}&8.46&9.67\%\\
		&\multirow{2}{*}{GD-MMA}&Signal &12.22&12.44&12.68&\\
		&&Noise &5.23&5.28&\textbf{4.82}&7.84\%\\
		\hline
		\multirow{4}{*}{Talbot}&\multirow{2}{*}{D-MMA}&Signal &8.22&8.13&8.35&\\
		&&Noise &3.49&\textbf{3.11}&3.52&10.89\%\\
		&\multirow{2}{*}{GD-MMA}&Signal &7.98&8.25&8.75&\\
		&&Noise &1.86&1.91&\textbf{1.68}&9.68\%\\
		\hline\hline
	\end{tabular}
\end{table}
~\\

\subsection{\textcolor{blue}{Real Experimental Results}}\label{sec:result3}
\begin{figure}[htbp]
	\centering
	\includegraphics[width=0.9\columnwidth]{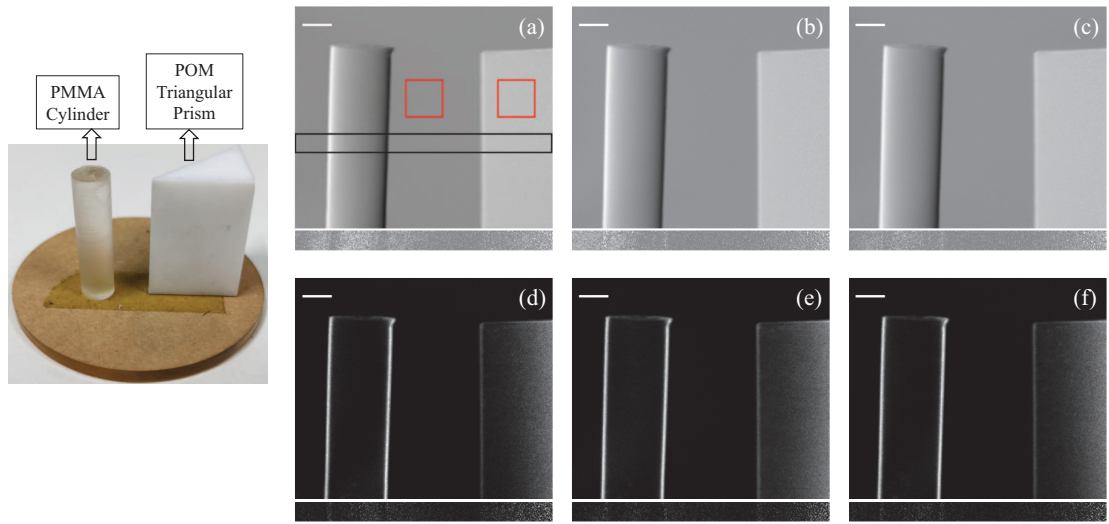}
	\caption{\label{fig:9}\textcolor{blue}{The used sample in the real experiment is shown on the left, while the DPC (a/b/c) and DFC (d/e/f) images retrieved by the D-MMA algorithm are shown on the right. Sub-figures (a/d) are acquired with the typical \textcolor{blue}{fluence} distribution, while (b/e) and (c/f) are acquired with the Type-1 and Type-2 adaptive fluence distribution, respectively. Their noise images by subtracting that of low dose data corresponding to the partial image in the black box (a) are displayed in the bottom of every sub-figure. The scale bar is 6 mm and the red box is the ROI area for evaluating the noise performance.}}
\end{figure}

\textcolor{blue}{As shown in Fig. \ref{fig:9}, we retrieved the DPC and DFC images from real experimental data by the D-MMA algorithm, which was acquired with the typical constant fluence and with our proposed two practical forms in Eq.(\ref{equ:16}) and Eq.(\ref{equ:17}), respectively. Since the total fluence is relatively high, all the retrieved images are of high image quality and low noise levels. However, on the subtracted images from the low dose images, it can be seen that the image noise with typical constant fluence distribution are higher than those of either the Type-1 or Type-2 practical adaptive fluence distributions. Due to the slightly mechanical instability of phase-stepping, there are some structure signals in the noise images.}

\textcolor{blue}{To compare the noises under different fluence distributions quantitatively, we chose two ROIs to evaluate the signals and noises as listed in Table \ref{tab:3}. ROI1 is the background area while ROI2 is the triangular prism sample area with a constant DPC signal. For the DPC image, it can be found that the noise reduction ratios range from 10\% to 15\%, similar with the synthetic experimental results. For the DFC image, the reduction ratios are also a little smaller than those of the DPC image, but the noise level is still lower than the typical constant fluence distribution.}

\begin{table}[htbp]
	\centering
	\caption{\label{tab:3}\textcolor{blue}{The real experimental signals and noise variances of DPC images($10^{-6}$) and DFC images ($10^{-12}$) in Fig. \ref{fig:9} retrieved by D-MMA under the typical constant and the two forms of adaptive fluence distributions.}}
	\begin{tabular}{cccccccc}
	\hline\hline
	Contrast & Area & Index & Typical & Type-1  & Redu-1 & Type-2& Redu-2 \\
	\hline
	\multirow{4}{*}{DPC}&\multirow{2}{*}{ROI 1}&Signal &1.72 &1.82&&1.73&\\
	&&Noise &1.55&1.32& 14.83\%&1.34&13.55\%\\
	&\multirow{2}{*}{ROI 2}&Signal &60.81&61.73& &61.63&\\
	&&Noise &2.88&2.52& 12.50\%&2.57&10.76\%\\
	\hline
	\multirow{4}{*}{DFC}&\multirow{2}{*}{ROI 1}&Signal &1.56&1.60& &1.64&\\
	&&Noise &1.61&1.40&13.04\%&1.42&11.80\%\\
	&\multirow{2}{*}{ROI 2}&Signal &21.74&21.98& &21.88&\\
	&&Noise &6.15&5.37&12.68\%& 5.65&8.13\%\\
	\hline\hline
	\end{tabular}
\end{table}
~\\

\section{DISCUSSION}\label{sec:discu}
From the theoretical derivations in Section \ref{sec:methods} and experimental validations in Section \ref{sec:results}, we demonstrated the effectiveness of our proposed \textcolor{blue}{fluence} adaptation strategy for task-based dose optimization in GBI and EI. This strategy of adjusting \textcolor{blue}{fluence} distributions in the PS process is inspired by the weighting kernel functions in the formulas of the D-MMA and GD-MMA algorithms, which tells us that the data from different PS steps may has different contributions to the noise in the retrieved contrasts. Starting from the unified formulas in the D-MMA and GD-MMA algorithms, we solve a optimization problem for noise variance minimization under the same total \textcolor{blue}{fluence}. Combining the common models of real original sample PSCs under a constant \textcolor{blue}{fluence}, i.e. the cosine function for GBI systems and the Gaussian function for EI systems, we can simplify the original optimization problem as shown in Section \ref{sec:methods} and obtain the ideal solution in Eq. (\ref{equ:160}), which is proportional to the absolute weighting kernel functions and the root of the original sample PSC. \textcolor{blue}{This conclusion in Eq. (\ref{equ:160}) is a key of our work, which provides a totally new fluence design in GBI and EI. The kernel functions is available after determining the target contrast and information retrieval algorithm, while the original sample PSC may be obtained using a low-dose pre-scanning scheme. Such prior images are commonly used in the traditional CT. They could become available for XPCI when it goes to wide applications in the future.}

Since the original sample PSC is sometimes unavailable in advance \textcolor{blue}{at present, we also propose two more practical adaptive fluence forms in Eq. (\ref{equ:16}) and Eq. (\ref{equ:17}). The Type-1 form approximates the sample PSC with the flat field PSC} as long as the overall phase shift by the scanning sample is small, which can be used in both GBI and EI systems. In particular for GBI systems, due to the \textcolor{blue}{relatively smaller weighting difference caused by the sample PSC (see about 25\%) compared with that by the kernel functions (up to 200\%),} the optimal adaptive \textcolor{blue}{fluence} distribution can further be approximated as the Type-2 form in Eq. (\ref{equ:17}). According to Eq. (\ref{equ:16}) and Eq. (\ref{equ:17}), we can design the \textcolor{blue}{fluence} adaptation mechanism specifically for a target contrast and an information retrieval algorithm in GBI and EI systems. In fact, non-uniform \textcolor{blue}{fluence} distribution may be also useful for other information retrieval methods, such as FCA, DB-MMA and TA-MMA\cite{wu2019truncated}, which is being investigated currently in our laboratory as an independent work. The proposed adaptive \textcolor{blue}{fluence} method can be directly used to improve the noise performance of the contrasts retrieved by the D-MMA and GD-MMA algorithms. Vice versa, if one wants to obtain contrasts with similar noise levels, the patient dose by using the adaptive \textcolor{blue}{fluence} method can be lower than that of the typical constant \textcolor{blue}{fluence} method. 

The simulated results in Section \ref{sec:result1} not only demonstrate the effectiveness of our proposed method for noise reduction, but also give out the effective ranges of the parameters in PSCs, such as the visibility (about $V=a_1/a_0<70\%$) in GBI systems, the standard deviation (about $\sigma<\pi/4$) and mean value (about $-\pi/8<\mu<\pi/8$) in EI systems. According to the literature, these parameter ranges are consistent with most real systems and can be used as a guidance to the use of our proposed adaptive method in practice.

In this paper, we have not taken \textcolor{blue}{much consideration} of the dark-field contrast mainly for two reasons. As seen in Eq. (\ref{equ:G3}), \textcolor{blue}{the dark-field contrast retrieved by MMA includes two parts, i.e. the second order normalized moment and the first-order normalized moment, which is related to the phase contrast. Therefore, the} first reason is that noises of phase contrast will affect noises of dark-field for both D-MMA and GD-MMA algorithms. The other one is that our proposed method is task-based and we can only adopt an optimal \textcolor{blue}{fluence} distribution for one contrast at a time. However, the way to get the optimal solution for the dark-field is similar with that for the phase-contrast according to our deviations in Section. \ref{sec:method1}, and therefore, it is \textcolor{blue}{not discussed in detail here}.

It is worth noting some limitations exist in our work. First and foremost, the proposed \textcolor{blue}{fluence} adaptation method has an underlying condition that the optimal \textcolor{blue}{fluence} distribution is the same for all pixels on the detector. Therefore, from our derivations in Section. \ref{sec:method1}, it can be known that the acquired \textcolor{blue}{flat field PSCs} at all pixels need to have a similar phase offset $\phi$ when the moving grating is at the same position. To satisfy this prerequisite, the system must have high perfection in grating manufacturing and grating assembly, especially in adjusting the grating parallelism. It is difficult to be perfect for the entire gratings and therefore, \textcolor{blue}{one generally have to choose some ROI regions to do the optimization. However, with the development of dynamically modulated x-ray sources, it may be able to generate intensity modulation for ROI regions.} Secondly, this method is task-based and can be only optimal for a specific contrast either the phase-contrast or the dark-field. It means that we can reduce the noise in one contrast, while it may increase the noise in the others. In the future, we will evaluate these effects for the non-target contrast and will try to balance the two independent \textcolor{blue}{fluence} distributions.

\section{CONCLUSION}\label{sec:concl}
In this work, based on recently developed analytic information retrieval algorithms and the \textcolor{blue}{Gaussian} noise model, we derive a \textcolor{blue}{fluence} adaptation mechanism for task-based dose optimization in x-ray phase-contrast imaging, which challenges the constant \textcolor{blue}{fluence} concept during the phase-stepping process. The simulated and experimental results validate our theory, with the noise variance of the phase-contrast being reduced \textcolor{blue}{10\% - 25\% under the same total absorption dose}. This work explores a new direction for dose reduction in x-ray phase-contrast imaging, especially for grating-based imaging and edge-illumination systems. Future work includes more studies for dose reduction on different x-ray phase-contrast imaging systems and different information retrieval algorithms using the proposed \textcolor{blue}{fluence} adaptation strategy.

\section*{ACKNOWLEDGMENTS}
This work was supported in part by the National Natural Science Foundation of China (No. 61527807, No. 81771829 and No. 62031020). And this work was carried out with the support of Shanghai Synchrotron Radiation Facility (SSRF). We want to thank Dr. Zhentian Wang and Dr. Carolina Arboleda for sharing the experimental data in SSRF, and thank Dr. Shengping Wang and Prof. Weijun Peng for offering the breast specimens from Fudan University Shanghai Cancer Center.

\section*{CONFLICTS OF INTEREST}
The authors have no relevant conflicts of interest to disclose.

\section*{\textcolor{blue}{Appendix A. The simplification of the object function}}\label{sec:app1}
\textcolor{blue}{To simplify the optimization task, we can assume the noise term $n_s(i)$ in Eq. (\ref{equ:3}) obeys the normal distribution $\mathcal{N}(0, M\cdot\bar{s}_0(i)t(i))$ according to the known statistical rules of the scintillator detectors\cite{G2000Radiation}, where $M$ is a constant factor close to 1.  Considering the signal term $\bar{s}_0(i)t(i)$ is much larger than the noise term $n_s(i)$, it is reasonable to conclude that the sum of signals is also much larger than the sum of noises, i.e., $\sum_{i=1}^N \bar{s}_0(i)t(i) \gg \sum_{i=1}^N n_s(i)$.} 

For GBI systems, as demonstrated in numerous literature \cite{momose2003demonstration, pfeiffer2006phase}, it is common to express the original PSC $\bar{s}_0(i)$ acquired with the constant \textcolor{blue}{fluence} distribution by using a cosinusoidal model, i.e., 
\begin{equation}
\label{equ:5}
\textcolor{blue}{\bar{s}_0(i)=a_0+a_1 \cos\left(\phi_i+\varphi_1^s\right), \phi_i=\frac{2\pi}{N}*(i-\frac{N+1}{2}) .}
\end{equation}
Then the denominator term in Eq. (\ref{equ:3}) can be calculated as, 
\begin{equation}
\label{equ:6}
\textcolor{blue}{\sum_{i=1}^N \left[\bar{s}_0(i)t(i)+n_s(i)\right]/t(i) \approx \sum_{i=1}^N \bar{s}_0(i)= Na_0.}
\end{equation}

For EI systems, the original sample PSC $s_0(i)$ can be modeled with a Gaussian function, i.e.,
\begin{equation}
\label{equ:18}\textcolor{blue}{ \bar{s}_{0}(i)=\frac{A}{\sqrt{2 \sigma^{2}}} \exp \left[-\frac{(\phi_i-\varphi^s)^{2}}{2 \sigma^{2}}\right], \phi_i=\frac{2\pi}{N}*(i-\frac{N+1}{2}),}
\end{equation}
where, $A$, $\textcolor{blue}{\varphi^s}$ and $\sigma$ represent the attenuation, phase-contrast, and dark-field information, respectively. \textcolor{blue}{As it is known that $\int_{-\infty}^{\infty} e^{-x^2}dx=\sqrt{\pi}$, the denominator in Eq. (\ref{equ:3}) can also be calculated as a constant, i.e.,}
\begin{equation}
\label{equ:19}\textcolor{blue}{ \sum_{i=1}^N \left[\bar{s}_0(i)t(i)+n_s(i)\right]/t(i) \approx \sum_{i=1}^N \bar{s}_0(i)= \sum_{n=1}^{N} \frac{A}{\sqrt{2 \sigma^{2}}} \exp \left[-\frac{(\phi_i-\varphi^s)^{2}}{2 \sigma^{2}}\right]=\sqrt{\pi}A,}
\end{equation}
\textcolor{blue}{which is largely valid for any $N\geq3$, and therefore it is required to acquire three images at least in the phase-stepping process.}

Therefore,  for both GBI and EI systems, the object function in Eq. (\ref{equ:3}) can be simplified as:
\begin{eqnarray}
L(t) &&= \text{Var}\left(\sum_{i=1}^N h(i)\left[\bar{s}_0(i)t(i)+n_s(i)\right]/t(i)\right) \nonumber\\
\label{equ:7} 
&&=\text{Var}\left(\sum_{i=1}^N h(i)\bar{s}_0(i)\right)+\text{Var}\left(\sum_{i=1}^N \frac{h(i)}{t(i)}n_s(i)\right)\\
\label{equ:8} &&=M\cdot\sum_{i=1}^N \frac{h^{2}(i)\bar{s}_0(i)}{t(i)}.
\end{eqnarray}
where, \textcolor{blue}{the first term about $\bar{s}_0(i)$ in Eq.(\ref{equ:7}) can be considered as noise-free according to our definition in Eq.(\ref{equ:1}), and the variance of the noise term $n_s(i)$ can be calculated as $M\cdot\bar{s}_0(i)t(i)$ according to the mentioned normal distribution of detector signals above.} 

\section*{\textcolor{blue}{Appendix B. An error analysis of the practical forms}}\label{sec:app2}
\textcolor{blue}{In Section \ref{sec:method3}, we proposed two practical forms of adaptive fluence distributions, which can applied directly without the original sample PSCs. To demonstrate the rationality of these approximation, we carried out an error analysis here.}

\textcolor{blue}{In order to compare results with the optimal form, one can calculate the optimal value of the object function in Eq. (\ref{equ:10}) by substituting the optimal distribution in Eq. (\ref{equ:160}), i.e.,
\begin{eqnarray}
\label{equ:B1}L(t^*)&&=\sum_{i=1}^N \frac{h^{2}(i)\bar{s}_0(i)}{|h(i)|\sqrt{\bar{s}_0(i)}}\cdot \frac{\sum_{i=1}^N |h(i)|\sqrt{\bar{s}_0(i)}}{N} \nonumber\\
&&=\frac{1}{N}\left[\sum_{i=1}^N |h(i)|\sqrt{\bar{s}_0(i)}\right]^2 \nonumber\\
&& = \frac{1}{N}\sum_{i=1}^N \sum_{j=1}^N  |h(i)h(j)|\sqrt{\bar{s}_0(i)\bar{s}_0(j)} 
\end{eqnarray} 
where, $\bar{s}_0(i)$ and $\bar{s}_0(j)$ denote the intensity of a pixel at $i$-th and $j$-th steps in the phase-stepping, respectively; $h_0(i)$ and $h_0(j)$ are also the corresponding kernel function weights.}

\textcolor{blue}{Then if we also take the first approximate form in Eq. (\ref{equ:16}) into the object function, the value can be written as follows:
\begin{eqnarray}
\label{equ:B2}L(t^*_1)&&=\sum_{i=1}^N \frac{h^{2}(i)\bar{s}_0(i)}{|h(i)|\sqrt{\bar{f}_0(i)}}\cdot \frac{\sum_{i=1}^N |h(i)|\sqrt{\bar{f}_0(i)}}{N}  \nonumber\\
&&= \frac{1}{N}\sum_{i=1}^N |h(i)|\frac{\bar{s}_0(i)}{\sqrt{\bar{f}_0(i)}} \cdot \sum_{i=1}^N |h(i)|\sqrt{\bar{f}_0(i)} \nonumber\\
&&=\frac{1}{N}\sum_{i=1}^N \sum_{j=1}^N  |h(i)h(j)|\bar{s}_0(i)\sqrt{\frac{\bar{f}_0(j)}{\bar{f}_0(i)}} \nonumber\\
&&= \frac{1}{N}\sum_{i=1}^N \sum_{j=1}^N  |h(i)h(j)|\sqrt{\bar{s}_0(i)\bar{s}_0(j)} \sqrt{\frac{\bar{s}_0(i)}{\bar{s}_0(j)}\cdot \frac{\bar{f}_0(j)}{\bar{f}_0(i)}}
\end{eqnarray} 
By subtracting Eq. (\ref{equ:B2}) from Eq. (\ref{equ:B1}), the absolute error can be obtained as:
\begin{equation}
\label{equ:B3}\epsilon_1(L) = L(t^*) - L(t^*_1) = \frac{1}{N}\sum_{i=1}^N \sum_{j=1}^N  |h(i)h(j)|\sqrt{\bar{s}_0(i)\bar{s}_0(j)} \left[1-\sqrt{\frac{\bar{s}_0(i)}{\bar{s}_0(j)}\cdot \frac{\bar{f}_0(j)}{\bar{f}_0(i)}}\right]
\end{equation} }

\textcolor{blue}{Similarly, for the second practical form, one can calculate the object function value as:
\begin{eqnarray}
\label{equ:B4}L(t^*_2)&&=\sum_{i=1}^N \frac{h^{2}(i)\bar{s}_0(i)}{|h(i)|}\cdot \frac{\sum_{i=1}^N |h(i)|}{N}  \nonumber\\
&&= \frac{1}{N}\sum_{i=1}^N \sum_{j=1}^N  |h(i)h(j)|\sqrt{\bar{s}_0(i)\bar{s}_0(j)} \sqrt{\frac{\bar{s}_0(i)}{\bar{s}_0(j)}}
\end{eqnarray} 
And the absolute error can be written as:
\begin{equation}
\label{equ:B5}\epsilon_2(L) = L(t^*) - L(t^*_2) = \frac{1}{N}\sum_{i=1}^N \sum_{j=1}^N  |h(i)h(j)|\sqrt{\bar{s}_0(i)\bar{s}_0(j)} \left[1-\sqrt{\frac{\bar{s}_0(i)}{\bar{s}_0(j)}}\right]
\end{equation} }

\textcolor{blue}{Taking the sample PSC expression of GBI in Eq.(\ref{equ:5}) as an example, one can express the ratio of  sample PSCs and flat field PSCs as:
\begin{eqnarray}
\label{equ:B6}&&\sqrt{\frac{\bar{s}_0(i)}{\bar{s}_0(j)}}=\sqrt{\frac{a_0^s+a_1^s \cos(\phi_i+\varphi_1^s)}{a_0^s+a_1^s \cos(\phi_j+\varphi_1^s)}} = \sqrt{\frac{1+V^s \cos(\phi_i+\varphi_1^s)}{1+V^s \cos(\phi_j+\varphi_1^s)}} \\
&&\sqrt{\frac{\bar{s}_0(i)}{\bar{s}_0(j)}\cdot\frac{\bar{f}_0(j)}{\bar{f}_0(i)}}=\sqrt{\frac{1+V^s \cos(\phi_i+\varphi_1^s)}{1+V^s \cos(\phi_j+\varphi_1^s)}\cdot  \frac{1+V^f \cos(\phi_i+\varphi_1^f)}{1+V^f \cos(\phi_j+\varphi_1^f)}}
\end{eqnarray} 
where $V^s, V^f$ are the visibilities of sample PSCs and flat field PSCs, respectively, both of which are usually smaller than 30\%. Therefore, the object function values of both two practical forms are close to that of the optimal form, and all these values are smaller than that of the typical constant distribution. }

\textcolor{blue}{Substituting $t_0(i)=1$ into Eq. (\ref{equ:10}) , we can also calculate the value of the object function under the typical constant distribution as:
\begin{equation}
\label{equ:B4}L(t_0)=\sum_{i=1}^N h^{2}(i)\bar{s}_0(i)
\end{equation} }

\begin{figure}[htbp]
	\centering
	\includegraphics[width=0.9\columnwidth]{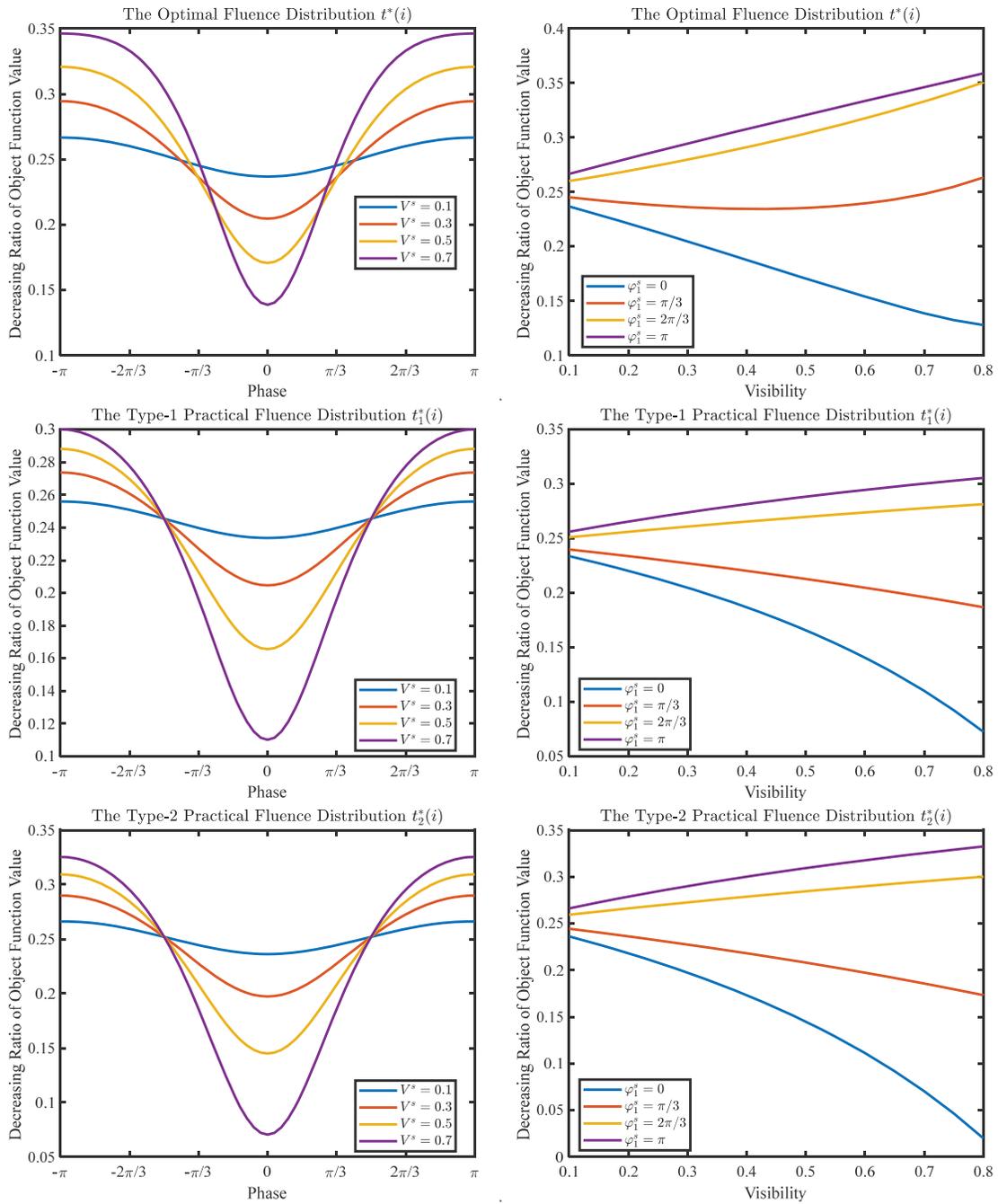}
	\caption{\label{fig:B1}\textcolor{blue}{The relative decreasing ratios between our proposed three distributions (including $t^*(i)$, $t_1^*(i)$ and $t_2^*(i)$) and the typical one ($t_0(i)$,), i.e. $1-\frac{L(t^*)}{L(t_0)}$.}}
\end{figure}

\textcolor{blue}{Based on the cosine expression of sample PSCs in Eq.(\ref{equ:5}), we change the visibility
$V^s$ and phase term $\varphi_1^s$ to assess the values of the object function under all distribution above (including $t_0(i)$, $t^*(i)$, $t_1^*(i)$ and $t_2^*(i)$). In Fig. \ref{fig:B1}, we plotted the relative decreasing ratios between our proposed three distributions and the typical one, i.e. $1-\frac{L(t^*)}{L(t_0)}$. From these results, it can be seen that all these distributions are better than the typical constant one, and the Type-1 form is better than the Type-2 form in terms of the noise variance.}

\section*{REFERENCES}
\addcontentsline{toc}{section}{\numberline{}References}
\vspace*{-20mm}



\bibliography{myref}      




\end{document}